\documentstyle[11pt,aaspp4,flushrt]{article}

\newcommand\gsim{\mathrel{\raise.3ex\hbox{$>$}\mkern-14mu
             \lower0.6ex\hbox{$\sim$}}}
\newcommand\lsim{\mathrel{\raise.3ex\hbox{$<$}\mkern-14mu
             \lower0.6ex\hbox{$\sim$}}}

\begin{document}

\title{STRUCTURAL AND DYNAMICAL UNCERTAINTIES IN MODELING AXISYMMETRIC ELLIPTICAL GALAXIES}
\author{Aaron J. Romanowsky \\
  Christopher S. Kochanek}
\affil{Harvard-Smithsonian Center for Astrophysics, MS-10, \\
       60 Garden Street,
       Cambridge MA 02138 \\
       Email: aromanowsky@cfa.harvard.edu}
\authoraddr{MS-10 \\ 
       60 Garden Street \\
       Cambridge MA 02138  \protect \\
       Email: aromanowsky@cfa.harvard.edu}
 
\begin{abstract}
Quantitative dynamical models of galaxies require deprojecting the observed surface 
brightness to determine the luminosity density of the galaxy.  Existing deprojection
methods for axisymmetric galaxies assume that a unique deprojection exists for any
given inclination, even though the projected density is known to be degenerate to
the addition of ``konus densities'' that are invisible in projection.  We develop
a deprojection method based on linear regularization that can explore the range of 
luminosity densities statistically consistent with an observed surface
brightness distribution.  The luminosity density is poorly constrained
at modest inclinations ($i\gsim 30^\circ$),
even in the limit of vanishing observational errors.
In constant mass-to-light ratio, axisymmetric, two-integral 
dynamical models, the uncertainties in the luminosity density result in large
uncertainties in the meridional plane velocities.
However, the projected line-of-sight velocities show variations
comparable to current typical observational uncertainties.
\end{abstract}

\section{INTRODUCTION}

Models of elliptical galaxies seek to understand the spatial distribution of
their stars (e.g. Franx, Illingworth, \& de Zeeuw \markcite{FID}1991; Statler \markcite{TS3}1995),
the structure of their orbits (Dehnen \& Gerhard \markcite{DG1}1993, \markcite{DG2}1994; Arnold, Robijn, \& de Zeeuw \markcite{ARD}1995), and to secure evidence
for the presence of dark halos (Saglia et al. \markcite{RS1}1993; Carollo et al. \markcite{CC1}1995)
or black holes (van der Marel et al. \markcite{RV2}1994; Dehnen \markcite{WD1}1995).
These problems can be probed using the observed velocities of the stars (van der Marel \markcite{RV1}1991; Carollo \& Danziger \markcite{CD1}1994), X-ray emission
(Fabbiano \markcite{GF1}1989, \markcite{GF2}1995),
gaseous disks and rings (Schweizer, Whitmore, \& Rubin \markcite{SWR}1983; Whitmore, McElroy, \& Schweizer \markcite{WM S}1987; Bertola et al. \markcite{FB1}1991; Franx, van Gorkom, \& de Zeeuw \markcite{FVD}1994),
and gravitational lenses (Maoz \& Rix \markcite{MR1}1993; Kochanek \markcite{CSK1}1995, \markcite{CSK2}1996).
Most modern dynamical models employ axisymmetric, two-integral models
in which the distribution function depends only on the energy and the angular
momentum about the symmetry axis (e.g. Binney, Davies, \& Illingworth \markcite{BDI1}1990; Hunter \& Qian \markcite{HQ1}1993).  Such models
have simple solutions to the Jeans equations (Satoh \markcite{CS1}1980; see also Binney \& Tremaine 1987 \S 4.2), and in some cases the distribution
function can be determined
(Dehnen \& Gerhard 1993, 1994; Qian et al. \markcite{EQ1}1995).
Quantitative application of the two-integral
models to a real galaxy relies on the deprojection of the observed surface brightness
of the galaxy to determine its three-dimensional luminosity density. 

The uniqueness of the deprojection of a galaxy depends on the symmetries of the
density.  Spherical galaxies always have unique deprojections, and
ellipsoidal galaxies with fixed axes have unique deprojections for 
known inclinations (e.g. Binney 1985).  General axisymmetric galaxies
do not have unique deprojections, except when the symmetry axis is in the plane 
of the sky (inclination $i=90^\circ$, Rybicki \markcite{GR1}1987).
The projection operation destroys information about the Fourier components of
the density which lie in a ``cone of ignorance'' of opening angle $90^\circ - i$ 
about the symmetry axis.  A ``konus density'' (Gerhard \& Binney \markcite{GB1}1996),
whose Fourier transform is non-zero only inside the cone of ignorance, can be
added to the luminosity density without changing the projected surface
brightness.  Kochanek \& Rybicki \markcite{KR1} (1996) developed methods
to produce families of konus densities with arbitrary equatorial density
distributions.  Simple konus densities generally look like ``disks'' because of the conical
symmetry of the model in Fourier space; this is consistent with Rix \& White's
\markcite{RW1}(1990) observation that disks would be nearly invisible in
ellipticals, unless close to edge-on.

We need to explore two issues to understand the effects of the deprojection degeneracy
on axisymmetric models of ellipticals.  The first issue is the existence of smooth
konus densities.  A real galaxy must have a smooth, monotonic, positive definite
density profile.  These physical restrictions limit the allowed range of
konus densities.  For example, none of the analytic konus densities of 
Gerhard \& Binney  \markcite{GB1}(1996) or Kochanek \& Rybicki \markcite{KR1} (1996)
are physical
because they have unacceptable angular profiles at large radii
due to the conditions imposed to find analytic solutions.
The second issue is that
even if the luminosity density is underconstrained, the degeneracy is dynamically
interesting only if the inferred velocity dispersions or velocity dispersion profiles
are uncertain by more than the observational errors.

The projection operator $\bf P$ is a linear operator $I = {\bf P } \nu$ between
the luminosity density $\nu$ and the surface brightness $I$.  Deprojection
corresponds to inverting the linear operator $\nu = {\bf P}^{-1} I$.  
Methods to invert $\bf P$ must cope with three generic problems: true
degeneracies, limited sampling, and amplification of noise. True degeneracies,
such as the konus densities, correspond to eigenvalues of $\bf P$ that are
zero, and they can be handled only by adding additional constraints
on the inversion such as smoothness, positivity, or monotonicity.  
The surface brightness is always discretely sampled, but we want to 
determine the continuous luminosity density.  If we try to determine
the luminosity density at more points than are sampled in the surface
brightness, the luminosity density is underconstrained by the data
and additional constraints are required to perform the inversion.    
Most inversions are also ill-conditioned, because some of the eigenvalues
of $\bf P$ are nearly zero.  Small eigenvalues of $\bf P$ are
large eigenvalues of ${\bf P}^{-1}$, and they amplify small fluctuations
of $I$ into large fluctuations in $\nu$.  The inversion method must
suppress these oscillations. 

Previous approaches to deprojection produce ``unique'' inversions because
they have additional, hidden constraints.  Lucy's \markcite{BL1}(1974) method 
is the standard deprojection algorithm used in stellar dynamical models 
(e.g. Newton \& Binney \markcite{NB1}1987; Binney et al. 1990;
van der Marel, Binney, \& Davies \markcite{VBD1}1990; Gerhard \markcite{OB1}1991; van der 
Marel 1991; Dehnen \markcite{WD1}1995).
It is a simple, iterative scheme that converges toward a
density distribution that exactly fits the data;
a ``unique'' solution results because of an implicit, nonparametric bias.
Lucy's method also introduces
numerical instabilities into the solution, so
the iterations are manually halted at some point when the density still ``looks good''.
Bendinelli's \markcite{OB1}(1991) method expands the surface brightness in terms of
Gaussian profiles, and then numerically fits the density to these terms.
Palmer's \markcite{PP1}(1994) method finds a finite angular polynomial series for the density
whose projection fits the observed surface brightness to arbitrary accuracy.
Functional fitting methods are fast, and frequently easy to program, 
but they are limited to a subset of the possible solutions.

None of these existing methods provides a way of studying the true degeneracies
of deprojection and their effects on dynamical models of axisymmetric galaxies.
In \S 2, we develop a deprojection method based on linear regularization that
allows us to explore the degeneracies of the inversion while keeping the
density well-defined and physical. 
In \S 3, we test the algorithm on artificial galaxy images.
In \S 4, we give results for two galaxies (NGC 1439 and NGC 7619), and in
\S 5 we summarize our conclusions.
Two appendices add some details of the numerical algorithms.

\vskip 11pt

\section{ METHODS}

\subsection{ Projection Geometry and Algorithm}
An axisymmetric galaxy has density distribution $\nu (X,Y,Z)$, where
the $Z$-axis is the symmetry axis of the galaxy. 
The surface brightness is $I(x,y)$, 
where $(x,y)$
are the coordinates in the plane of the sky, and the $z$-axis 
lies along the line of sight. The $X$- and $x$-axes coincide,
and
correspond to the line of nodes at the intersection of the plane of the
sky and the equatorial plane of the galaxy (thus, for an oblate
axisymmetric galaxy, the $x$-axis corresponds to the observed major
axis of the galaxy).
Because the galaxy is axisymmetric, its density distribution is
completely specified by $\nu(R,Z)$,
where $R=\sqrt{X^2+Y^2}$.
For simplicity, we assume that the galaxy has a reflection
symmetry about the equatorial plane, $\nu(Z)=\nu(-Z)$.
The surface density is related to the luminosity density by the projection operation,
\begin{equation}
I(x,y) = \int_{-\infty}^{+\infty} \nu (R,Z) dz.
\end{equation}
The transformation between the galaxy's
coordinates and the observer's coordinates is
\begin{eqnarray}
 X &=& x \nonumber \\
 Y &=& y \cos i - z\sin i \nonumber \\ 
 Z &=& y \sin i + z\cos i  \nonumber \\
 r^2 = R^2 + Z^2 &=& \varpi^2 + z^2 , 
\end{eqnarray}
where $\varpi^2 \equiv x^2 + y^2$,
and the inclination $i=90^\circ$ corresponds to an ``edge-on'' galaxy with no cone
of ignorance.

The density of the galaxy $\nu(R,Z)$ is divided into zones,
$\nu_{jk} \equiv \nu(\theta_j,r_k)$, where one quadrant of the
meridional plane $(R,Z)$ is divided into
$N_r$ radial zones and into $N_a$ angular zones.
The radial zones are logarithmically spaced in $r$ from inside $\varpi_{\rm min}$ to outside $\varpi_{\rm max}$,
where $\varpi_{\rm min}$ and $\varpi_{\rm max}$ are the radii of the inner
and outer surface density measurements.
The angle $\theta$ is the standard spherical polar angle
with $\theta=0$ on the symmetry axis; the angular zones are
equally spaced in $\theta$ ($\theta_j = j \cdot \Delta \theta$, where $j=[0\cdots N_a-1]$ and $\Delta \theta = \pi/2N_a$).
Reflection symmetry about the equatorial plane is implicit in the model,
although the symmetry could be trivially removed.

Although the zones $\nu_{jk}$ represent the real density,
the projection algorithm considers the galaxy to be made of ``stacked
blocks'' in order to force the density to vary monotonically in angle\footnote{A general
two-dimensional density distribution that is monotonic in both angle
and radius cannot be constructed in this way.}. 
For an oblate galaxy, the total density of each successive zone increases with its
angle $\theta_j$, and each additional contribution can be thought of as a block of density
$\check{\nu}_{jk}=p_{jk}^2$ stacked on top of the previous blocks, with total density
\begin{equation}
\nu_{jk} = \sum_{i=0}^{i=j} \check{\nu}_{ik} = \sum_{i=0}^{i=j} p_{ik}^2.
\end{equation}
By construction, a density model expressed in terms of $p_{jk}$ is positive definite,
oblate\footnote{For a prolate galaxy, the density decreases with angle $\theta_j$: $\nu_{jk} = \sum_{i=j}^{N_a-1} \check{\nu}_{ik}$.}, and monotonic in angle.

The projected intensity is sampled at discrete points $I_{lm}=I(x_{lm},y_{lm})$, where
$l,m$ are arbitrary indices (e.g. corresponding to polar coordinates $\varpi_l$, $\psi_m$).
The contribution to $I_{lm}$ 
from density block $\check{\nu}_{jk}$ is found by integrating along
the line of sight through the block
\begin{equation}
dI_{jklm} = \int_{z_1}^{z_2} \check{\nu}_{jk} (z) dz ,
\end{equation}
where $(z_1,r_1)$, $(z_2,r_2)$  are the coordinates of the intersection
points of the line of sight
from $(x_{lm},y_{lm})$ with the edges of zone $\{jk\}$, and
\begin{equation}
I_{lm} = \sum_{jk} dI_{jklm}.
\end{equation}
We use a first-order projection scheme that linearly interpolates the density
between radially-adjacent zones with
\begin{equation}
\check{\nu}_{jk} (z) = \check{\nu}_{jk} + \frac{\check{\nu}_{jk+1} - \check{\nu}_{jk}}{r_k - r_{k+1}} [r_k - r(z)] ,
\end{equation}
giving the first-order approximation
\begin{equation}
dI_{jklm} = \frac{z_2-z_1}{r_{k+1}-r_k}(\check{\nu}_{jk}r_{k+1}-\check{\nu}_{jk+1}r_k) + \frac{1}{2} \frac{\check{\nu}_{jk+1}-\check{\nu}_{jk}}{r_{k+1}-r_k}\left[z_2 r_2 - z_1 r_1 + \varpi^2 \ln \left(\frac{z_2+r_2}{z_1+r_1}\right)\right]_{jklm} .
\end{equation}
(see Appendix A). Beyond the outermost density zones, the density is assumed to decrease
as a power law, normalized to the density of the outermost zone at a given angle $\theta_j$:  $\nu \propto \nu_{j0} (r^2+s_{\rm b}^2)^{-\alpha_{\rm b}/2}$.
The (small) contribution to the projection from
this density ``tail'' is found by numerical integration of equation (4).

\subsection{Smoothing and Regularization}

Given the projection, $I_{lm}$, of the current model density distribution,
and the observed surface brightness, $I^*_{lm}$, we define
a $\chi^2$ statistic for the goodness of fit, 
\begin{equation}
\chi^2 = \sum_{lm} \left(\frac{I_{lm}-I^*_{lm}}{\sigma_{lm}}\right)^2, 
\end{equation}
where $\sigma_{lm}$ is the noise associated with the measurement
$I^*_{lm}$ (we neglect any correlation function of the noise). If there are $N_{\rm data}$ points $I_{lm}$, then a good fit should have
$\chi^2 \simeq N_{\rm data}$ with a one standard deviation error of
$\Delta\chi^2\equiv\chi^2-N_{\rm data}\simeq \pm\sqrt{2N_{\rm data}}$ (for $N_{\rm data} \gg 1$) if we neglect the
number of degrees of freedom in the source model. The algorithm used to
minimize the function requires the first and second
derivatives of the $\chi^2$  with respect to the density kernels
$p_{lm}$; these derivatives are given in Appendix B.

Linear regularization is used to combat
the problems that occur in unregularized solutions of
integral equations such as equation (1). The primary problem is the
tendency for the density to oscillate between the radial sampling points
when there are more model density points than surface sampling 
points. The simplest solution is to find a good fit to the data while 
simultaneously minimizing a smoothing function
\begin{equation}
H_1 = \sum_{jk}\left(\frac{h_{jk+1}\nu_{jk+1} - h_{jk}\nu_{jk}}{h_{jk} \nu_{jk}}\right)^2, 
\end{equation}
defined by the fractional variation in the radial density profile,
weighted by the bias function $h_{jk}\equiv h(r_k,\theta_j)$.
The bias function is used to weight the smoothness
equally over all points, and to provide a bias slope at large radii
where sky-level uncertainties poorly constrain the radial profile.
The density should also vary 
smoothly with angle, so we add a second smoothing function
\begin{equation}
H_2 = \sum_{jk} \left(\frac{h_{j+1k}\nu_{j+1k}-h_{jk}\nu_{jk}}{h_{jk}\nu_{jk}}\right)^2.
\end{equation}
to prevent unrealistically sharp angular density variations.
Ideally, $h(r,\theta)$ would be given self-consistently by the 
deprojected radial profile $\nu^{-1} (r,\theta)$. In practice, we fit
a simple analytic model, e.g. the power-law model
\begin{equation}
I(\varpi) = I_0 (s_{\rm b}^2 + \varpi^2)^{(1-\alpha_{\rm b})/2},
\end{equation}
to the surface data points along the major axis,
and analytically deproject it to give $h(r)$; then we assume an constant axis ratio
$q_{\rm b} \equiv b/a$
to give the angular variation $h(r_k,\theta_j)$, e.g.,
\begin{equation}
h_{jk} = \left[s^2 + r^2_k(\sin^2\theta_j+\frac{1}{q_{\rm b}^2}\cos^2\theta_j)\right]^\frac{\alpha}{2}.
\end{equation}
Such radial surface brightness models qualitatively cover much of the observed range of galactic
morphologies (Binney \& Tremaine \markcite{BT1}1987).
In addition to enforcing smoothness, the function $h(r,\theta)$ allows us
to explicitly bias the profile to some arbitrary morphology.

We then minimize the function $F = \chi^2 + \lambda H$, where
$H = H_1 + \kappa H_2$. The weighting factor $\kappa$ is somewhat arbitrary,
but must be small enough to keep the radial profile acceptably smooth.
For each image $I^*$ and bias function $h$, $\kappa$ is set by
trial-and-error to be as large as is possible before
significant radial fluctuations appear in the solution.
The Lagrange multiplier $\lambda$ must be
adjusted so that when $F$ is minimized, the 
$\chi^2$ is found to have a value in the range $N_{\rm data} \pm \sqrt{2 N_{\rm data}}$, where $N_{\rm data}$
is the number of surface density sampling points
(we used solutions with $\chi^2$ in the range $N_{\rm data} \pm \sqrt{N_{\rm data}}$).
The Lagrangian multiplier $\lambda$ is found
iteratively: an initial $\lambda$ is chosen, the function is minimized,
and the resulting $\chi^2$ value is linearly interpolated to $\chi^2=N_{\rm data}$ to
predict the correct $\lambda$; then the function is again minimized with the
new $\lambda$, and so on, until the value of the $\chi^2$ falls within the proper range.
The minimization is performed using the
Polak-Ribiere conjugate gradient method (Press et al. \markcite{WP1}1992).
The expressions for the gradients of $\chi^2$, $H_1$, and $H_2$
are given in Appendix B. 

\subsection{Velocity Calculations}

To examine the dynamical effects of the projection degeneracies we assume a constant mass-to-light
ratio, axisymmetric, two-integral dynamical model (e.g. Binney \& Tremaine 1987, van der Marel 1991).
Following Binney et al. (1990),
we calculate the potential $\Phi$ from the mass density $\rho = \Upsilon_0 \nu$ by first expanding the 
density in Legendre polynomials,
\begin{equation}
\rho_l(r') = \Upsilon_0 \int_0^\pi \nu (r',\theta) {\rm P}_l(\cos{\theta}) \sin{\theta}d\theta
\end{equation}
and then by finding the potential produced by the Legendre expansion of the density
\begin{equation}
\Phi(r,\theta) = -2\pi G \sum_l {\rm P}_l (\cos{\theta})\left[ \frac{1}{r^{(l+1)}}\int_0^r \rho_l(r') r'^{(
l+2)} dr'  + r^l \int_r^\infty \rho_l(r') \frac{dr'}{r'^{(l-1)}} \right].
\end{equation}
The solutions to the Jeans equations for the two-integral model are
\begin{equation}
\nu \sigma^2(R,Z) = \int_{Z}^\infty \nu (R,Z') \frac{\partial \Phi (R,Z')}{\partial Z'} dZ' ,
\end{equation}
and 
\begin{equation}
\langle v_\phi^2 \rangle (R,Z) = \sigma^2 + R \frac{\partial \Phi}{\partial R} + \frac{R}{\nu}\frac{\partial (\nu \sigma^2)}{\partial R} ,
\end{equation}
where the galaxy is assumed axisymmetric, steady-state ($\langle v_Z \rangle = \langle v_R \rangle = 0$), 
and isotropic in the meridional direction ($\sigma_R=\sigma_Z\equiv \sigma$) 
(see Binney \& Tremaine 1987 \S 4.2).  The projected line-of-sight velocity dispersion is 
\begin{eqnarray}
I \langle v^2_{\rm los} \rangle (x,y) &=& \int_{-\infty}^\infty dz \left[\nu \sigma^2 (R,Z) (\cos^2i +\sin^2\phi \sin^2 i) + \nu \langle v_\phi^2 \rangle \cos^2\phi \sin^2i \right] . \\
 &=& \int_{-\infty}^\infty \nu \sigma^2 (R,Z) dz + \sin^2i\int_{-\infty}^{\infty} R \cos^2\phi \left[\nu \frac{\partial \Phi}{\partial R} + \frac{\partial (\nu \sigma^2)}{\partial R} \right] dz.
\end{eqnarray}
We numerically integrate equations (15) and (16) to find the
velocity components $\sigma^2$ and $\langle v^2_\phi \rangle$ in the galaxy's meridional plane,
and then we numerically integrate equation (17) to find $\langle v_{\rm los}^2 \rangle$
(we do not separate out its components $\langle v_{\rm los}^2\rangle = \sigma_{\rm los}^2 + \langle v_{\rm los}\rangle^2$).
The accuracy of the numerical integration algorithm was verified with the analytic solution
of Satoh (1980; eqs. [8] and [13]) and other more {\it ad hoc} analytic solutions.

\section{ TEST PROBLEMS}

We test the method in four stages.  First we confirm that if we bias the solution toward the
true density, the method converges to the correct result.  Next we demonstrate that the 
deprojection is degenerate by examining the range of solutions we can produce by altering
the bias function.  To show that the degeneracy is not a numerical artifact, we next investigate 
the effects of the numerical resolution, observational errors, and numerical errors.
Finally we examine the dynamical effects of the degeneracy on the two-integral models.

\subsection{Deprojection with Correct Bias}

Before examining real galaxies, we studied the deprojection
algorithm and the effects of degeneracy with artificial images
of known density distributions.
We used the axisymmetric, oblate ($q_0 < 1$) density distribution,
\begin{equation}
\nu(R,Z) = \nu_0 \left(1 + \frac{R^2}{s_0^2} + \frac{Z^2}{s_0^2}\frac{1}{q_0^2}\right)^{-\alpha_0/2},
\end{equation}
which has the analytically calculable surface brightness,
\begin{equation}
I(x,y) = \frac{\pi^{1/2}\Gamma(\frac{\alpha_0-1}{2})}{\Gamma(\frac{\alpha_0}{2})}\frac{\nu_0 q_0 s_0 }{\sqrt{q_0^2\sin^2i+\cos^2i}}\left(1+\frac{x^2}{s_0^2}+\frac{y^2}{s_0^2}\frac{1}{q_0^2\sin^2i+\cos^2i}\right)^{(1-\alpha_0)/2}.
\end{equation}
In the test image, the surface brightness points $I_{lm}\equiv I(\varpi_l,\psi_m)$ are 
logarithmically-spaced in radius $\varpi$ (46 points from $\varpi /s_0 = 0.53$ to $\varpi /s_0 = 38.2$), 
and equally-spaced in angle $\psi$ (7 points from $\psi = 0$ to $\psi = \pi/2$).

The errors and uncertainties in the image of a bright galaxy are dominated by systematic effects rather than by photon
counting noise.  Imperfect flat-fielding produces a fractional error which varies over large scales 
in an image, and the sky background level is uncertain to within some constant value.
For a set of image points $I^*_{lm}$, we adopted an error model with $\sigma_{lm} = \sigma_0 I^*_{lm} + I_{\rm b}$.
To simulate a real image, we added artificial errors to our ``data''.
A constant background
offset, $I_{\rm b}$, simulated an error in determining the sky background level, and a Gaussian
distribution of fractional errors (with rms amplitude $\sigma_0$, and Fourier width corresponding to fluctuation wavelengths of about 5\% the full-frame width of the image) was added to simulate flat-fielding errors.
These model errors
qualitatively resemble the residuals from fitting real images.

We deprojected simulated images for various values of
the model parameters ($\alpha_0$, $s_0$, $q_0$, $i$), the biasing parameters
($\alpha_{\rm b}$, $s_{\rm b}$, $q_{\rm b}$), and error levels ($\sigma_0$, $I_{\rm b}$).
The results of some of these deprojections are shown in Figure 1,
where the image has the analytic form specified by equation (20), with
$\alpha_0 = 3.0$, $s_0 = 1.7$, $q_0 = 0.6$, $\sigma_0 = 0.01$ (corresponding
to a 0.01 magnitude error), $I_{\rm b} \simeq 0.0004 I^*_{\rm max}$ (sky level underestimated), 
and $i = 30^\circ$ (recall that $i=0^\circ$ is pole-on), with simulated errors added as described above.
The first column of Figure 1 shows the case in which we set the bias
function equal to the ``correct'' density: $h=\nu_{\rm corr}^{-1}$ (i.e.,
$\alpha_{\rm b}=\alpha_0$, $s_{\rm b}=s_0$, $q_{\rm b}=q_0$).
In this test
case, the final solution almost exactly matches the input model, with
rms fractional errors of: $\langle(\nu-\nu_{\rm corr})/\nu_{\rm corr}\rangle_{\rm rms} = 0.009$.
Similar deprojections were performed over a wide range of model parameters, with similar results:
when given the proper bias functions, the algorithm
reconstructed the original density function.
Large changes in the artificial error parameters
had negligible effects.

\subsection{The Existence of Degeneracies}

Our real interest, however, is how the results can differ from the
input density model given the degeneracies in the projection operator.
Figure 1 also shows three other deprojections of the same image,
where each column shows a solution with a different bias function $h$.
We deliberately chose extreme biases and a low inclination
($i=30^\circ$) to highlight the range of possible solutions.
In the second column, the bias function is more elliptical than the ``correct''
solution ($\alpha_{\rm b} = 3.0$, $s_{\rm b} = 1.7$, $q_{\rm b} = 0.3$); in the third column it
decreases more rapidly with radius ($\alpha_{\rm b} = 3.5$, $s_{\rm b} = 2.0$, $q_{\rm b} = 0.6$);
and in the fourth column we have added an exponential disk
($h^{-1}= \nu_{\rm corr} + k_{\rm d} \exp(-R/R_{\rm b}) {\rm sech} (Z/Z_{\rm b})$, 
with $k_{\rm d}$ = 0.25, $R_{\rm b}/s_0=5.2$, and $Z_{\rm b}/s_0=0.9$).
Each model is converged until it is a statistically acceptable fit to
the data ($\chi^2 \simeq N_{\rm data}$).
The rms fractional differences in the density for the solutions are
$\langle(\nu-\nu_{\rm corr})/\nu_{\rm corr}\rangle_{\rm rms} =  0.58$, $0.35$, and $0.22$,
for columns 2, 3, and 4, respectively,
demonstrating that the konus densities can have large amplitudes.
These densities qualitatively resemble the disks examined by Rix \& White 
\markcite{RW1}(1990) and the analytic solutions of Gerhard \& Binney \markcite{GB1}(1996) and 
Kochanek \& Rybicki \markcite{KR1}(1996), without the
discontinuity problems exhibited by those solutions at large radii.
Although some of the solutions show systematic 
deviations in the outer regions of the galaxy at the two-sigma level, and the
density structure of the disk-biased solution has some unphysical features, 
it should be remembered that we deliberately chose the bias functions to find extreme examples.

\subsection{ Searching for ``True'' Konus Densities}

We next checked the effects on the solutions of varying the numerical resolution.
A true projection degeneracy should exist independently of the sampling
of either the image or the density distribution, except in the limit
that the data overconstrains the density.
Our standard $N_r \times N_a$ density grid has $N_r = 100$ radial 
and $N_a = 25$ angular zones.  There are numerical errors in the projection, and
if we compute the errors for the test problem in \S3.1, the contributions
to $\chi^2/N_{\rm data}$ from numerical errors are $\simeq 0.003$, $0.01$, and $0.07$ 
for $i=30^\circ$, $60^\circ$, and $90^\circ$ respectively, given
our standard error model.
Although these numerical
errors are much smaller than the observational errors ($\chi^2/N_{\rm data}\simeq 1$),
we must examine whether the magnitude of the degeneracies is influenced by the
resolution.

We performed a series of tests comparing the numerical projection of the
test model to the analytic projection of the test model, gradually reducing 
the density zone resolution.  Reduced resolution increases the numerical projection
errors and reduces the range of possible solutions because the number of degrees
of freedom in the density model ($N_r\times N_a$) approaches the number of data points
($N_{\rm data}=322$). The inclination was fixed at $i=30^\circ$. For zone
resolutions of $N_r \times N_a = 2500$, $1600$, $400$, $196$, and $100$, the resulting
numerical errors were $\chi^2_{\rm err}/N_{\rm data} \simeq 0.003$, $0.008$, $0.2$, $1.2$,
and $6.0$, respectively. The $\chi^2$ from numerical errors became unacceptable only
in the two lowest resolution simulations, where
the density distribution was overconstrained ($N_{\rm zones} \equiv N_r \times N_a < N_{\rm data}$).

Next we deprojected the test model ($q_0=0.6$) using substantially rounder
($q_{\rm b}=0.9$) or flatter ($q_{\rm b}=0.3$) bias densities.  As we reduced the resolution,
the allowed degeneracy towards low-ellipticity models was little affected,
but the degeneracy towards high-ellipticity models was substantially reduced
(see Figure 2).  The amplitude of the degeneracy did not change between our
standard resolution and the next lowest resolution.  The inability to produce
high ellipticity models at low numerical resolutions is due to the decreasing
angular resolution.  Note, however, that the degeneracies still exist even when
the data overconstrains the density, a clear sign that the degeneracies
are not due to the numerical resolution of the calculation.

A true degeneracy should also exist independently of the observational errors.
The amplitude of the degeneracy may increase with the amount of noise in
the observations, but it should not vanish in the limit of no noise.
We made a sequence of deprojections of a single image while gradually reducing the 
observational error levels, $\sigma_0$ and $I_{\rm b}$, until the numerical
errors become significant ($\chi^2_{\rm err}/N_{\rm data} \sim 0.1$); $I_{\rm b}$ was
reduced in proportion to $\sigma_0$.  Figure 3 shows the deprojections
of the same image used in Figure 1 ($\alpha_0=3.0, s_0=1.7, q_0=0.6$), biased to 
either $q_{\rm b}=0.9$ and $q_{\rm b}=0.3$ as a function of the inclination ($90^\circ$, $60^\circ$, 
or $30^\circ$), with varying errors.
The permitted range of the solutions depends strongly on the inclination
and weakly on the errors.
At $i=90^\circ$ the solution is unique, with any
uncertainty in the axis ratio due to the noise; the rms fractional density variation 
between the extreme solutions ($q_{\rm b}=0.9$, $0.3$) at ``normal'' noise levels ($\sigma_0 = 0.01$) is
$\langle (\nu_{0.3}-\nu_{0.9})/\nu_{0.9}\rangle_{\rm rms} = 0.26$.
At $i=60^\circ$ the axis ratio can be biased
upwards and downwards by $\Delta q \sim 0.1$ about the true axis
ratio; for $\sigma_0 = 0.01$, the allowed rms variation is
$\langle (\nu_{0.3}-\nu_{0.9})/\nu_{0.9}\rangle_{\rm rms} = 0.53$.
At $i=30^\circ$ the axis ratio can be biased by $\Delta q \sim 0.2$ 
about the true axis ratio. For $\sigma_0 = 0.01$, the allowed rms variation is
$\langle (\nu_{0.3}-\nu_{0.9})/\nu_{0.9}\rangle_{\rm rms} = 0.82$;
even when the errors are at the level of $0.002$ magnitudes, the allowed rms variation is
$\langle (\nu_{0.3}-\nu_{0.9})/\nu_{0.9}\rangle_{\rm rms} = 0.67$.

In all previous tests,
we fit our projections to the analytic surface brightness model,
so numerical projection errors limited how far we 
could reduce the noise.  For our final test we ``removed'' these errors
by using the numerical projection of
the density model as the image, so that as the model observational noise approaches
zero there is a numerical solution with $\chi^2/N_{\rm data}=0$.       
Figure 4 shows a sequence of deprojections of the same $q_0=0.6$ model,
with $i=30^\circ$,
in which the solution was biased to $q_{\rm b}=0.3$ and $q_{\rm b}=0.9$, and the
stated error level was gradually decreased from $\sigma_0=0.01$ to
$\sigma_0 = 10^{-4}$ (no artificial noise was added); for speed, a coarser
density grid was used ($N_r \times N_a = 52 \times 13$).
The allowed ellipticity range in the density
decreased as $\sigma_0$ decreased, but at a much slower rate than the
residuals in the image (Figure 5). 
For a true konus solution we should see the
rms variation of the density become constant and finite at zero noise, but the presence of any noise
always allows the variations to be larger. Gerhard \& Binney (1996) refer
to such additional degeneracies as ``truncated konus densities'', whose projections are never
exactly zero but have large rms density variations for very small projected
surface densities.  
The continued presence of a coherent 46\% rms density variation, even in
the limit that the errors are one hundred times smaller than typical
observational data, means that for all practical purposes we have found
true degeneracies of the projection operator rather than any numerical 
effect.

\subsection{ Dynamical Consequences}

The existence of a deprojection degeneracy is only of academic importance unless
it has dynamical consequences.  The fundamental question is whether the dynamical
uncertainties introduced by the deprojection uncertainties can alter the conclusions
of constant mass-to-light ratio, axisymmetric, two-integral dynamical models
(e.g.  van der Marel 1991).  For a fixed mass-to-light ratio,
we calculated the velocities (as described in \S 2.3) in the meridional plane, $\sigma^2$ and 
$\langle v_\phi^2 \rangle$, and the projected mean square line-of-sight velocity $\langle v_{\rm los}^2 \rangle$
for solutions to the analytic galaxy image ($q_0=0.6$); these velocity profiles are
shown in Figure 6 as a function of inclination and bias function.
The total luminosity (and thus the total mass) of the solutions is fixed,
and the average radial distribution of the luminosity cannot vary a great deal,
so the total meridional plane velocities $\sqrt{2\sigma^2+\langle v_\phi^2\rangle}$
(and thus the kinetic energy)
remain essentially constant, with variations typically $\lsim$20\% at $30^\circ$
and $\lsim$3\% at $90^\circ$;
this is a consequence of the virial theorem.
However, the velocity anisotropy $\langle v_\phi^2\rangle^{1/2}/\sigma$
varies considerably,
with larger variations at lower inclinations
($\lsim$60\% at $30^\circ$ vs. $\lsim$25\% at $90^\circ$).
Projection of these velocities results in weak variations
in the line-of-sight velocities,
taking the form of alterations
in the ratio between the major and minor axis profiles.
Typical variations in $\langle v_{\rm los}^2 \rangle^{1/2}$ at $i=90^\circ$ are $\lsim$10\%, at $i=60^\circ$ are $\lsim$20\%,
and at $i=30^\circ$ are $\lsim$30\%.
Such variations are roughly comparable to typical measurement errors in line-of-sight velocities.

\section{REAL EXAMPLES: NGC 7619 AND NGC 1439}

For our final experiment we selected for deprojection two galaxies
from van der Marel (1991),
NGC 7619 and NGC 1439; these galaxies were chosen for their small
isophote twists.
The photometric profiles were taken
from Franx, Illingworth, \& Heckman \markcite{FIHa}(1989a, hereafter FIHa), where the 
R-band photometry is parametrized by an ``intermediate'' axis ($m \equiv \sqrt{ab}$)
profile for each galaxy, along with the ellipticity,
isophote twist, and higher-order isophote corrections at each radial point. 
We converted the data into a series of radial profiles and their errors
at 7 evenly spaced angles beginning on the major axis and ending on the minor axis. 
We incorporated the isophote twist and $\cos 4\psi$ terms into our data,
but none of the other higher order terms
(strictly speaking, the isophote twist, which is an indicator of triaxiality,
violates our density symmetry requirements, but in practice its presence
was of little consequence for the final solutions).
We made no corrections for the seeing, so the density profiles inside $\sim 5$
arcseconds will be unreliable.  The photometric uncertainties are 
dominated by flat-fielding errors (at the 1\% level) except at the outer radii,
where there is a sky level uncertainty of 1\%-3\%.  Both galaxies were deprojected 
for a variety of inclinations and biasing parameters.

The kinematic data were taken
from Jedrzejewski \& Schechter \markcite{JS1}(1989, hereafter JS)
and from Franx, Illingworth, \& Heckman \markcite{FIHb}(1989b, hereafter FIHb),
where the velocity dispersion and rotation velocity are given along the major and
minor axes.
The line-of-sight velocities were then derived from the density solution as in \S 3.4, and fit to the kinematic
data (by minimizing the $\chi^2$ between them) to find the $R$-band mass-to-light ratio $\Upsilon_R$; 
to minimize the effects of seeing on the results,
and to allow for direct comparison with van der Marel (1991),
any velocity data point inside
4 arcsec is not used in the fit.
Several of these solutions
are shown in Figures 7 and 8.

NGC 7619 is an E2/E3 elliptical with a small ($<13^\circ$)
isophote twist; van der Marel \markcite{RM1}(1991) was not able to fit an
acceptable axisymmetric two-integral dynamical model for this galaxy. The
projected ellipticity varies from 0.16 to 0.28, with small ($<0.7\%$) higher-order
corrections. As is evident from the residuals in Figure 7, there is
a radial ``kink'' in the outer regions of the galaxy ($r \sim$ 50 arcsec) which cannot be
well fit by a single power-law.  The deprojections can be made modestly
more or less elliptical, with some solutions showing boxy and
``S0--like'' structures.
The range of the solutions is
not as large as found for the test galaxies (compare Figure 7 to Figure 3),
probably because much of the ``smoothness'' was taken up in trying to fit to a 
single radial power-law. 
The rms fractional density variations between the extreme
solutions ($q_{\rm b}=0.9$ and $q_{\rm b}=0.5$) for inclinations of $90^\circ$,
$60^\circ$, and $40^\circ$ are
$\langle(\nu_{0.5}-\nu_{0.9})/\nu_{0.9} \rangle_{\rm rms}$ = 0.07, 0.15, and 0.34, respectively
($i=40^\circ$ was used because a convergence difficulty caused minimization time to be prohibitive for $i=30^\circ$).
For $i=90^\circ$, the solutions (with different biases $q_{\rm b}$)
had $R$-band mass-to-light ratios of $(\Upsilon/\Upsilon_\odot)_R \simeq$ (3.9$\pm$0.2)$h_{50}$ and $\chi^2/N \sim$ 2.8 for the kinematic fit, where $H_0= 50 h_{50}$ km s$^{-1}$ Mpc$^{-1}$;
the error bars represent the range of solutions with $\Delta \chi^2 = \pm 4$
(where the poor fits of these solutions were renormalized to set $\chi^2=N$).
For $i=60^\circ$, the range of solutions had $(\Upsilon/\Upsilon_\odot)_R \simeq$ (3.8-3.9$\pm$0.3)$h_{50}$ and $\chi^2/N \sim$ 3.2-4.3. 
For $i=40^\circ$, the solutions had $(\Upsilon/\Upsilon_\odot)_R \simeq$ (4.1$\pm$0.3)-(4.1$\pm$0.4)$h_{50}$ and $\chi^2/N \sim$ 3.4-5.8. 
Given the uncertainties and the fact that neither we nor van der Marel found an
acceptable ($\chi^2/N \sim 1$) solution for the velocities, our mass-to-light ratios $\Upsilon_R$ are consistent with
van der Marel's $(\Upsilon/\Upsilon_\odot)_R \simeq (4.0 \pm 0.1) h_{50}$ at $90^\circ$ and
$(\Upsilon/\Upsilon_\odot)_R \simeq (4.1 \pm 0.1) h_{50}$ at $60^\circ$.

The second galaxy, NGC 1439, is an E1 elliptical with a small twist angle $<9^\circ$
and an indication of a disklike distortion in the inner parts (FIHa).  The projected 
ellipticity varies from 0.07 to 0.11, with higher-order corrections $<0.8\%$.
It has a counter-rotating core (FIHb).
As can be seen in the residuals in Figure 8, the projected galaxy has a strong variation 
in ellipticity with radius which is not well fit by a constant ellipticity.
Solutions biased toward high ellipticities again develop disk-like structures
that sometimes show a feature along a line at the inclination angle from the symmetry axis.
The rms fractional density variations between the extreme
solutions ($q_{\rm b}=1.0$, $q_{\rm b}=0.6$) for inclinations of $90^\circ$,
$60^\circ$, and $30^\circ$ are
$\langle(\nu_{0.6}-\nu_{1.0})/\nu_{1.0} \rangle_{\rm rms}$ = 0.05, 0.11, and 0.33, respectively.
For $i=90^\circ$, the solutions had $(\Upsilon/\Upsilon_\odot)_R \simeq$ (2.3$\pm$0.3)$h_{50}$ and $\chi^2/N \sim$ 1.7. 
For $i=60^\circ$, the solutions had $(\Upsilon/\Upsilon_\odot)_R \simeq$ (2.3$\pm$0.3)$h_{50}$ and $\chi^2/N \sim$ 1.7-1.8. 
For $i=30^\circ$, the solutions had $(\Upsilon/\Upsilon_\odot)_R \simeq$ (2.4-2.5$\pm$0.3)$h_{50}$ and $\chi^2/N \sim$ 1.6-1.8. 
Given the uncertainties, our mass-to-light ratios are consistent with van der Marel's
$(\Upsilon/\Upsilon_\odot)_R \simeq (2.2 \pm 0.1) h_{50}$.

Both galaxies show similar variations of the line-of-sight velocities.
The velocities in the meridional plane show large variations ($\lsim$20\% for NGC 1439
at $i=30^\circ$),
creating differences in the projected velocities that are smaller ($\lsim$7\%) than typical measurement errors.
The same geometric effects which create the konus densities
appear to also create ``konus velocities'' whose large amplitudes in the
meridional plane practically vanish in projection;
thus, considerable refinement of spectroscopic techniques would be needed to
rule out any deprojection degeneracies with velocity measurements.
It is possible that higher-order velocity moments
are not as strongly affected by the konus degeneracy, 
but an analysis of this question was outside the scope of this paper.
Note that none of the solutions actually fits the velocity data well.
While this does not alter our conclusions about the degeneracy of a real deprojection
solution, 
an accurate model would need to account for seeing effects, a varying
mass-to-light ratio, anisotropies, and triaxiality, in order to acceptably fit the data.
Note also that only two-integral models were used, and it is possible that
three-integral axisymmetric models would show larger kinematic variation in projection.

\section{ CONCLUSIONS}

The deprojection of an axisymmetric galaxy is uniquely specified only
if $i=90^\circ$ and the symmetry axis is in the plane of the sky.  At all
other inclinations there is a gradually increasing degeneracy in the projection
operator corresponding to an unconstrained ``cone of ignorance'' with
opening angle $90^\circ - i$ in the 
Fourier transform of the density (Rybicki 1987).  Recent
analytic studies by Gerhard \& Binney (1995) and Kochanek \& Rybicki (1996)
have found simple density functions, called konus densities, that are
invisible in projection because their Fourier transforms are non-zero only
in the cone of ignorance.

We have developed a new deprojection method based on linear
regularization and explored the effects of the deprojection degeneracy on the inferred
structure and dynamics of axisymmetric elliptical galaxies.  The advantages of our approach 
over earlier methods are that it is non-parametric, that it performs a well-defined statistical 
fit to the surface brightness data, that it strictly enforces the positivity and monotonicity of the
solution, and that it allows us to explore the degeneracies of the projection
operator.  The standard method of Lucy (1974) is non-parametric but does not
have a well-defined convergence criterion, and functional fitting methods 
such as Palmer's (1994) and Bendinelli's (1991) depend on parametric forms.
No previous method has been able to explore the degeneracies of the projection
operator, or to impose monotonicity or any analytic requirement on the models.

We find that axisymmetric galaxies have large deprojection uncertainties at modest
inclinations even when we are restricted to positive definite, monotonic
density distributions.  The uncertainties are not due to numerical projection
errors, insufficient grid resolution, or observational noise, although increasing the noise
in the observations increases the uncertainty.  Even when the observational
error in the surface brightness points approaches $10^{-4}$ mag, it is possible to have
density distributions fitting the data with rms fractional variations of 46\% for $i=30^\circ$.
The differences between the model densities are the konus densities, and they
resemble the analytic solutions found by  Gerhard \& Binney (1995) and Kochanek \& 
Rybicki (1996).  If the bias function used to produce variations in the 
model density is not too spherical or too elliptical compared to the true density,
the resulting model density looks reasonable.  Solutions biased toward very
high ellipticities show strong ``disk-like'' structures with a feature
at angle $i$ from the equatorial plane, and could be rejected as physical
inversions. 
Because of the noise and konus degeneracy,
the solutions are quite sensitive to the choice of the bias function;
we chose simple functions which do not reproduce well some of the more complicated
features in the data (e.g. NGC 7619's radial kink), but one could easily
implement a more ``accurate'' bias by using Lucy's method or a parameterized approximate
deprojection method to arrive at an initial bias function.

We have evaluated the dynamical variations allowed by the deprojection uncertainties
in the constant mass-to-light ratio, axisymmetric, two-integral dynamical model 
(Binney et al. 1990; van der Marel 1991).
Although the velocities in the meridional plane can have large variations, 
the variations in the projected mean square velocities are modest for
all inclinations when compared to typical measurement errors. 
Given our structural and dynamical assumptions,
current velocity measurements are not helpful in reducing the deprojection degeneracy ---
the konus densities are associated with what we might term ``konus velocities''.
We infer mass-to-light ratios for NGC 1439 and NGC 7619 which are comparable to
those of van der Marel (1991), but with larger uncertainties. None of the
constant mass-to-light models fits the data well, with typical $\chi^2/N_{\rm dof} \simeq$ 1.5-3.

The implications of our results for more complicated models are unclear.
Statler \markcite{TS1}(1994a, \markcite{TS2}1994b) and Statler \& Fry \markcite{SF1}(1994)
have used dynamical models to tightly constrain the deprojection of a
triaxial galaxy, but they have assumed axis-ratios constant with
radius, severely limiting the generality of the solution.
Merritt \markcite{DM1}(1996) has developed a technique for deriving a unique
two-integral distribution function from surface brightness and velocity moment
measurements, but only at the non-degenerate edge-on inclination;
presumably, the large uncertainty in the density due to the degeneracy must
also affect the inferred distribution function.

\vskip 1cm

ACKNOWLEDGMENTS

\noindent
We thank George Rybicki, Dan Fabricant, and Roeland van der Marel
for their helpful comments.

\newpage
\appendix
\centerline{\small APPENDICES}
\section{First-Order Projection Scheme}
From the interpolated density of equation (6), we find its contribution to the projection:
\begin{eqnarray}
dI_{jklm} &=& \int_{z_1}^{z_2} \left( \check{\nu}_{jk}+\frac{\check{\nu}_{jk+1}-\check{\nu}_{jk}}{r_k-r_{k+1}}r_k \right) dz - \int_{z_1}^{z_2} \frac{\check{\nu}_{jk+1}-\check{\nu}_{jk}}{r_k-r_{k+1}}r(z)dz \nonumber \\
&=& \left( \check{\nu}_{jk}+\frac{\check{\nu}_{jk+1}-\check{\nu}_{jk}}{r_k-r_{k+1}}r_k \right) (z_2-z_1) - \frac{\check{\nu}_{jk+1}-\check{\nu}_{jk}}{r_k-r_{k+1}}\int_{z_1}^{z_2} r(z)dz .
\end{eqnarray}
Since $r^2 = \varpi^2 + z^2$,
\begin{eqnarray}
\int r dz &=& \int_{z_1}^{z_2} (\varpi^2+z^2)^{1/2}dz \nonumber \\
&=& \frac{z}{2}\sqrt{\varpi^2+z^2}+\frac{\varpi^2}{2}\ln \left(z+\sqrt{\varpi^2+z^2}\right) \nonumber \\
&=& \frac{zr}{2} + \frac{\varpi^2}{2}\ln (z+r),
\end{eqnarray}
which leads to the first-order approximation in equation (7).

\section{Gradient Expressions}
The derivatives of the $\chi^2$ statistic are:
\begin{equation}
{\chi^2_{in}}' \equiv \frac{\partial \chi^2}{\partial p_{in}} = 2 p_{in} \sum_{lm}\frac{I_{lm}-I^*_{lm}}{\sigma_{lm}^2} A_{inlm},
\end{equation}
\begin{equation}
{\chi^2_{in}}{''} = \frac{{\chi^2_{in}}'}{p_{in}} + 2\check{\nu}_{in}\sum_{lm}\left(\frac{A_{inlm}}{\sigma_{lm}}\right)^2 ,
\end{equation}
where
\begin{eqnarray}
A_{inlm} \equiv& &\Delta z_{i-1nlm} - \Delta z_{inlm} \nonumber \\
&+&2\cdot\left(\frac{\Delta z_{inlm} r_{n+1}}{r_{n+1} - r_n} + \frac{\Delta z_{in-1lm} r_{n-1}}{r_{n-1} - r_n}\right) \nonumber \\
&+& \frac{1}{r_n-r_{n+1}}\left[z_2r_2-z_1r_1+\varpi^2_{lm}\ln\left(\frac{z_2+r_2}{z_1+r_1}\right)\right]_{inlm} \nonumber \\
&+& \frac{1}{r_n-r_{n-1}}\left[z_2r_2-z_1r_1+\varpi^2_{lm}\ln\left(\frac{z_2+r_2}{z_1+r_1}\right)\right]_{in-1lm}.
\end{eqnarray}
The derivatives of the radial and angular smoothness parameters $H_1$ and $H_2$ are:
\begin{eqnarray}
(H_1)_{in}' &=& 4 p_{in} \sum_{j=i}^{N_a-1} \left\{\begin{array}{ll}
\frac{h_{jn}}{h_{jn-1} \nu_{jn-1}}\left(\frac{h_{jn}\nu_{jn}}{h_{jn-1}\nu_{jn-1}}-1\right) , & n > 0 \\
\frac{h_{jn+1}\nu_{jn+1}}{h_{jn} \nu_{jn}^2}\left(1-\frac{h_{jn+1}\nu_{jn+1}}{h_{jn}\nu_{jn}}\right), & n < N_r - 1 
			\end{array}\right. \\
(H_1){''}_{in} &=& \frac{(H_2)_{in}'}{p_{in}} + 8\check{\nu}_{in}\sum_{j=i}^{N_a-1}\left\{\begin{array}{ll}
\frac{h_{jn}^2}{h_{jn-1}^2 \nu_{jn-1}^2} , &  n > 0 \\
\frac{h_{jn+1}\nu_{jn+1}}{h_{jn} \nu_{jn}^3}\left(\frac{3h_{jn+1}\nu_{jn+1}}{h_{jn} \nu_{jn}}-2\right) , & n < N_r - 1  
			\end{array}\right. 
\end{eqnarray}
\begin{eqnarray}
(H_2)_{in}' &=& 4p_{in}\times \left\{\displaystyle{\begin{array}{ll}
\frac{h_{in}}{h_{i-1n}\nu_{i-1n}}\left(\frac{h_{in}\nu_{in}}{h_{i-1n}\nu_{i-1n}}-1\right), & i > 0 \\
\sum_{j=l}^{N_a-2} \frac{h_{j+1n}\check{\nu}_{j+1n}}{h_{jn}\nu^2_{jn}}\left(1-\frac{h_{j+1n}\nu_{j+1n}}{h_{jn}\nu_{jn}}\right), & i < N_a -1 
			\end{array}}\right. \\
(H_2){''}_{in} &=& \frac{(H_2)_{in}'}{p_{in}} + 8\check{\nu}_{in} \times \left\{\begin{array}{ll}
\left(\frac{h_{in}}{h_{i-1n}\nu_{i-1n}}\right)^2, & i > 0 \\
\sum_{j=l}^{N_a-2} \frac{h_{j+1n}\check{\nu}_{j+1n}}{h_{jn}\nu^3_{jn}}\left[\frac{h_{j+1n}}{h_{jn}}\left(\frac{3\check{\nu}_{j+1n}}{\nu_{jn}}+2\right)-2\right], & i < N_a - 1 
			\end{array}\right.
\end{eqnarray}
\newpage

\vskip 9pt
\centerline{\small CAPTIONS}
\vskip 9pt

\figcaption{
Effects of the bias function on deprojections of an artificial
galaxy image.
Each column represents a different choice for the bias function.
The top row shows the normalized residuals of the solution, $(I-I^*)/\sigma$, where the
horizontal dotted lines are the one-standard-deviation errors.
The dashed lines are the residuals of the projection of the ``correct'' density
(i.e. they show the added ``noise''),
and the curves are offset for visibility with the major axis on top.
The second row shows the deprojected density profiles on fixed
azimuths.
The third row shows contour plots of the density solution (solid
contours) and the ``correct'' density (dotted contours).
There are 26 contour levels shown, logarithmically spaced from $\nu=0.25$ to $\nu=2.5\times 10^{-6}$.
The bottom row shows contour plots of the konus density, or the difference between the final solution and
the ``correct'' density; dotted contours represent negative values.
There are 42 contour levels shown (positive and negative), logarithmically spaced from $\nu=10^{-2}$ to $\nu=10^{-6}$.
}

\figcaption{
The effects of numerical resolution on deprojections of artificial
galaxy images.
The inclination is fixed at $i=30^\circ$.
The major-to-minor axis ratio is plotted for each solution as a function of radius.
The ``correct''
solution is the central dashed line with $q_0=0.6$; the upper and lower
sets of curves are the solutions when biased toward $q_{\rm b}=0.9$ and $q_{\rm b}=0.3$.
The solid, dashed, dotted, and dot-dashed curves show the solutions
when the density zone resolution is $N_r\times N_a=100\times 25$,
$80\times 20$, $40\times 10$, $28\times 7$, and $20\times 5$, respectively. Note that
the image resolution is $N_x\times N_y =46\times 7$, so the
density is overconstrained by the data in the last two cases.
}

\figcaption{
Deprojections of an artificial galaxy image as a function
of noise level and inclination.
Each column represents a different assumed inclination $i$, with
$i=90^\circ$, $60^\circ$, and $30^\circ$, from left to right.
In the two top rows, the solid, dotted, and dot-dashed lines show
solutions with decreasing fractional noise $\sigma_0$.
The highest noise level is set to a realistic value;
the lowest is the level at which numerical errors
become significant.
The top row shows the density axis ratio, where the central dashed line is the
deprojection using the ``correct'' bias.
The middle row shows the  density solution angular profile at a fixed radius ($r/s_0=1.6$).
The bottom row shows contours of the solution density. The dotted and dashed lines show
solutions at the highest (normal) noise level ($q_{\rm b}=0.9$ and $q_{\rm b}=0.3$,
respectively); the solid contours show the ``correct'' solution, $q_0=0.6$.
There are 26 contour levels shown, logarithmically spaced from $\nu=0.25$ to $\nu=2.5\times 10^{-6}$.
}

\figcaption{
Deprojection of an artificial galaxy image with numerical errors
removed (see text),
as a function of the noise level $\sigma_0$.
The inclination is fixed at $i=30^\circ$.
The axis ratio is plotted for each solution. The ``correct''
solution is the central dotted line with $q_0=0.6$; the upper and lower
sets of curves are the solutions when biased toward $q_{\rm b}=0.9$ and $q_{\rm b}=0.3$.
The solid curves show a sequence of solutions with noise level $\sigma_0$ lowered.
}

\figcaption{
Effects of observational errors on deprojection.
The rms fractional variation in the density solution and the rms fractional
error in its projected
image are shown as a function of error level. The dotted line solutions were
biased toward $q_{\rm b}=0.3$, and the solid line solutions were biased
toward $q_{\rm b}=0.9$.
}

\figcaption{
Velocity dispersions for an analytic test galaxy ($q_0=0.6$),
assuming a constant mass-to-light ratio.
The left column shows the velocities in the meridional plane, along the major axis
(the variations along the minor axis were inconsequential).
The right column shows the projected velocities along
the major (upper profiles) and minor (lower profiles) axes.
The top box shows solutions for $i=90^\circ$, the middle for $60^\circ$, and
the bottom for $30^\circ$. At each inclination, solutions are shown
with biases of $q_{\rm b}=0.9$, $0.6$, and $0.3$. The vertical axis has arbitrary
units.
}

\figcaption{
Deprojection solutions and resultant projected velocities for NGC 7619,
for varying inclination and bias function.
Each column represents a different assumed inclination $i$.
The top row shows the normalized residuals of the solution: $(I-I^*)/\sigma$, where the
horizontal dotted lines are the one-standard-deviation errors.
The middle row shows the rms line-of-sight velocity profile for the major
(upper) and minor (lower) axes. Velocity data (JS; FIHb) are superimposed,
where the crosses are major axis points, the solid squares are minor axis points,
and the error-bars are one-$\sigma$ errors. 
The bottom row shows contours of the solution density.
There are 28 contour levels, from $\nu=0.25$ to $\nu=10^{-6}$.
The solid lines indicate a solution with a $q_{\rm b}=0.7$ bias function; 
the dotted lines, $q_{\rm b}=0.9$; the dashed lines, $q_{\rm b}=0.5$.
The vertical dot-dash line indicates the cut-off radius (4 arc sec) below which
the velocity data are not used.
}

\figcaption{
Deprojection solutions and resultant projected velocities for NGC 1439,
for varying inclination and bias function.
Velocity data are from FIHb.
Labels are the same as in Fig. 7.
There are 24 contour levels, from $\nu=0.25$ to $\nu=6.3\times 10^{-6}$.
The solid lines indicate a solution with a $q_{\rm b}=0.8$ bias function; 
the dotted lines, $q_{\rm b}=0.99$; the dashed lines, $q_{\rm b}=0.6$.
}

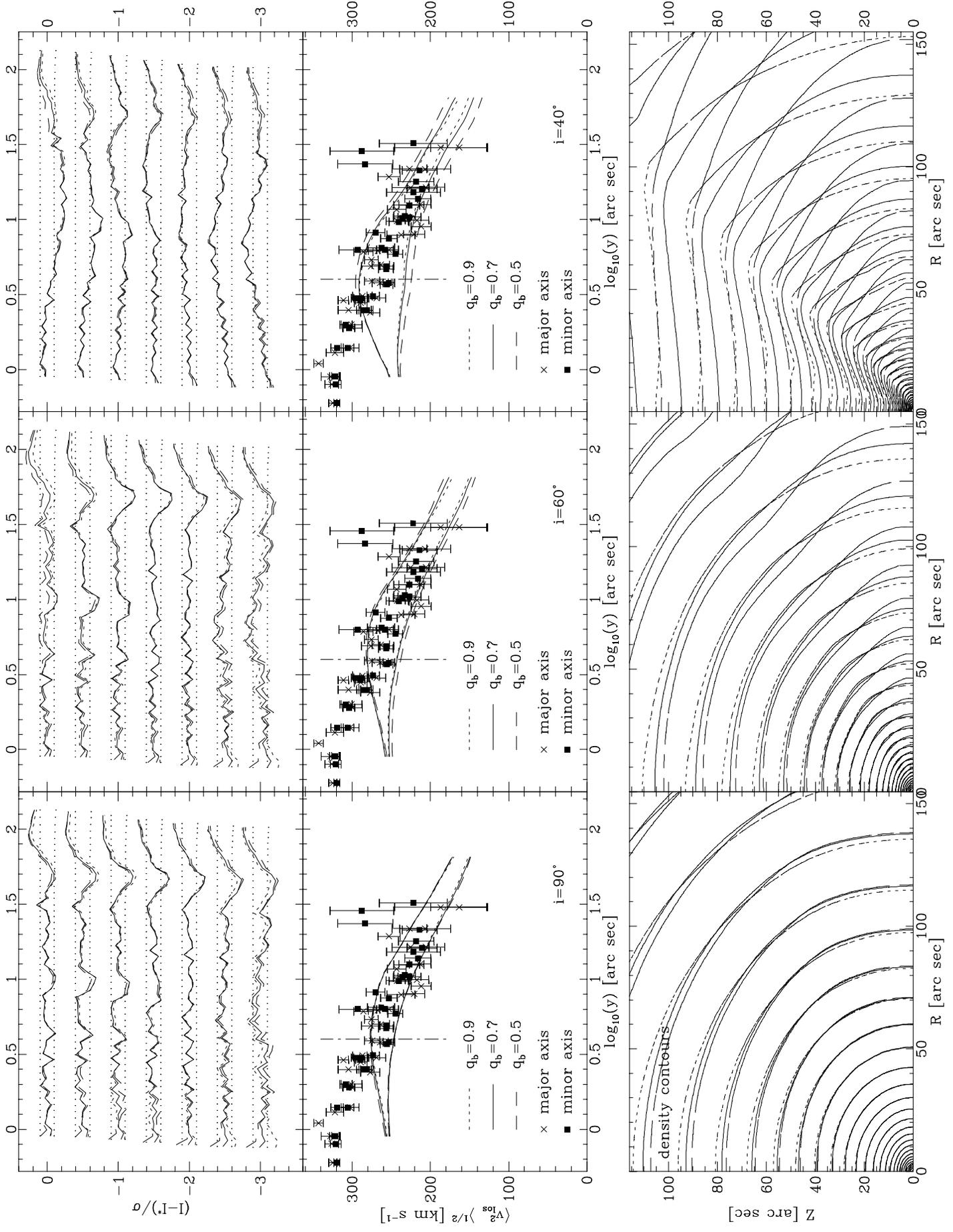
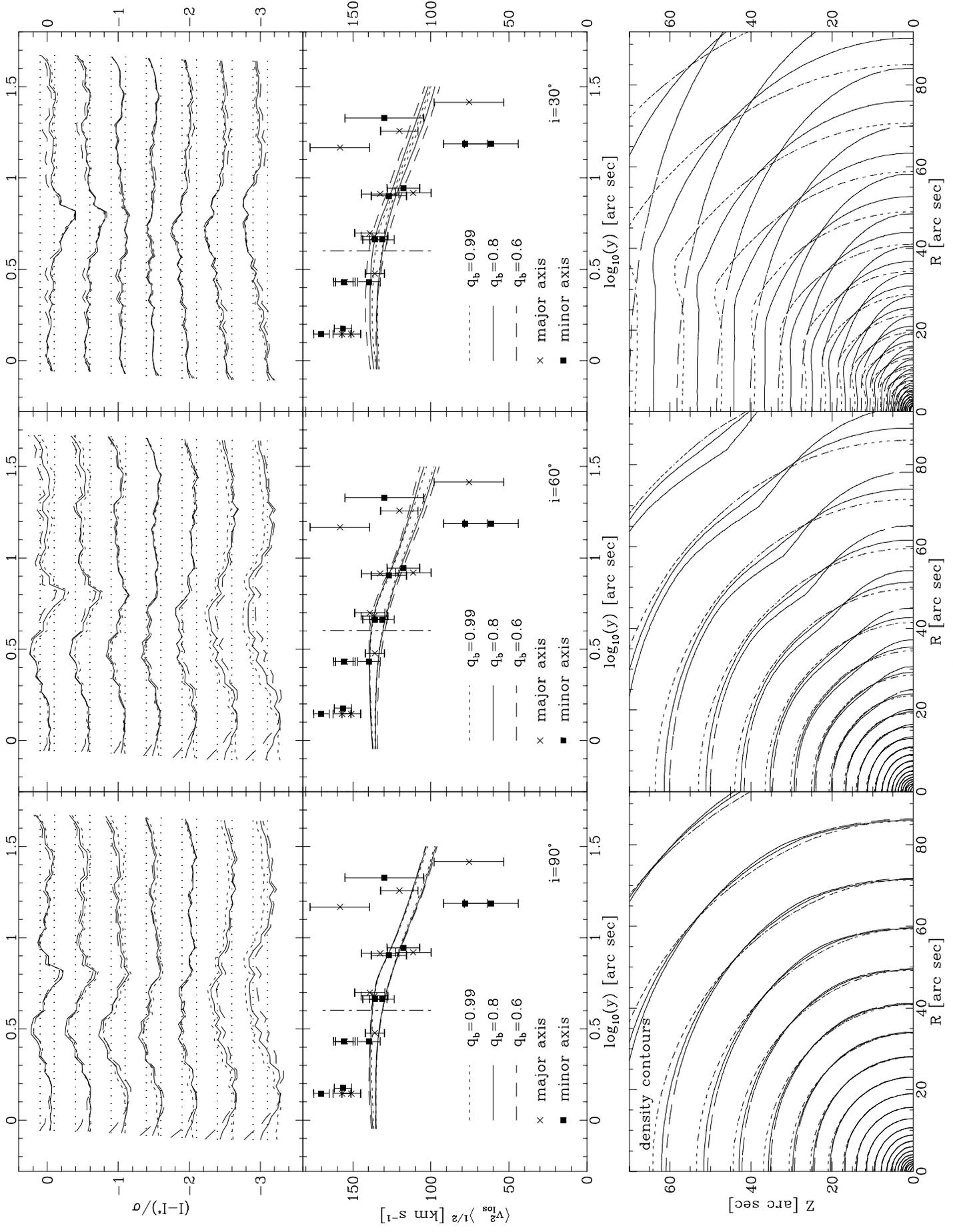
\end{document}